\documentclass[conference,twocolumn,10.5pt]{IEEEtran}
\usepackage{cite}
\usepackage{citesort}
\usepackage[top=0.705in, left=0.625in, bottom=1in, right=0.625in]{geometry}
\usepackage{amsmath}  
\usepackage{amssymb}
\usepackage{amsmath}
\usepackage{dsfont}
\usepackage{isomath}
\usepackage{amssymb}

\usepackage{array}
\usepackage{ifpdf}
\ifpdf
\usepackage[pdftex]{graphicx} 
\usepackage{epstopdf}         
\else
\usepackage{graphicx}         
\fi

\usepackage{varwidth}

\usepackage{mathtools}

\usepackage{algorithm}
\usepackage{algpseudocode}

\begin{document}

\title{{Delay-Efficient and Reliable Data Relaying  in Ultra Dense Networks using Rateless Codes}}
\author{\IEEEauthorblockN{Luyao Shang, Morteza Hashemi, Taejoon Kim, and Erik Perrins\\}
\IEEEauthorblockA{Department of Electrical Engineering \& Computer Science\\
University of Kansas, Lawrence, KS 66045\\
E-mails: \{lshang, mhashemi, taejoonkim, esp\}@ku.edu \\}
}

\maketitle

\begin{abstract}
We investigate the problem of delay-efficient and reliable data delivery in ultra-dense networks (UDNs) that constitute macro base stations (MBSs), small base stations (SBSs), and mobile users.  Considering a two-hop data delivery system, we propose a \emph{partial decode-and-forward (PDF)} relaying strategy together with a simple and intuitive amicable encoding scheme for rateless codes to significantly improve user experience in terms of end-to-end delay. Simulation results verify that our amicable encoding scheme is efficient in improving the intermediate performance of rateless codes. It also verifies that our proposed PDF significantly improves the performance of the decode-and-forward (DF) strategy, and that PDF is much more robust against channel degradation. Overall, the proposed strategy and encoding scheme are efficient towards delay-sensitive data delivery in the UDN scenarios.
\end{abstract}

\section{Introduction}
\label{sec:Intro}
Within the last few years, we have witnessed a dramatic growth in the number of connected wireless devices that generate huge amounts of mobile data traffic. According to the Cisco Visual Networking Index (VNI), the global mobile data traffic was 0.24 exabytes (EB) per month in 2010. This number has grown to 40.77 EB per month so far in 2020, nearly a 170-fold increase over the past decade. To better serve this explosive demand in wireless data, ultra-dense network (UDN) has become one of the key enabling technologies in the fifth generation (5G) systems since it significantly improves the overall throughput by deploying small cells near the users~\cite{Hoadley2012,Dhillon2012}. The small cells include relatively low-power base stations that cover geographical areas within a short distance from users. 

\begin{figure}[t]
	\centering\includegraphics[width=3.5in]{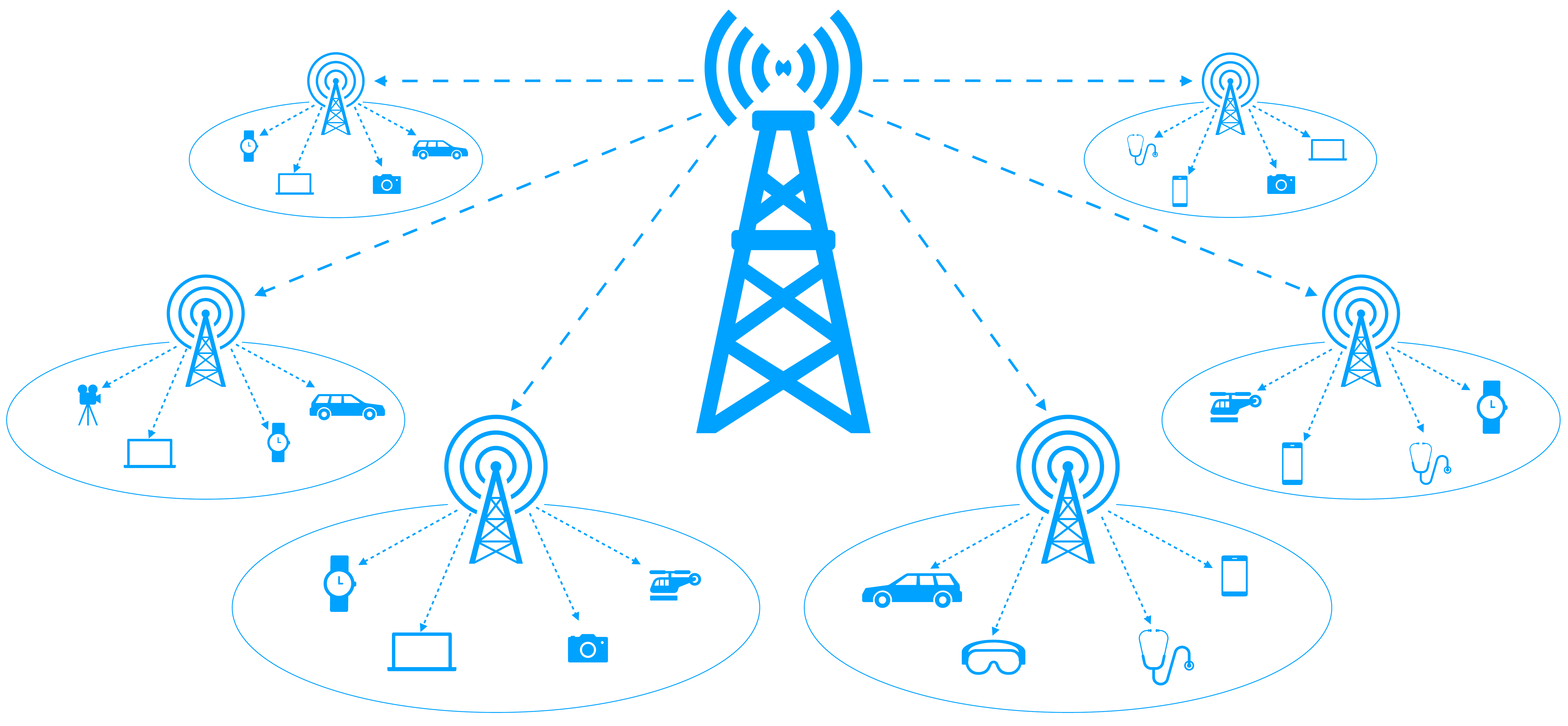}
	\caption{Illustration of the downlink communications of an ultra-dense network. The network includes a MBS and multiple SBSs with a number of users. The SBSs are densely deployed within a short distance from the users. The MBS fetches data from the content provider through the core network router, and delivers it to the users that are served by the SBSs via a two-hop link that operates on orthogonal spectra.}
	\label{fig:UDN}
\end{figure}

As shown in Fig.~\ref{fig:UDN}, the UDN constitutes the macro base station (MBS), small base stations (SBS), and users. The communications between MBS and users are supported by SBSs, forming a two-hop relay system. Popular relaying strategies such as decode-and-forward (DF) and amplify-and-forward (AF) are commonly adopted in multi-hop relay systems. The AF outperforms the DF in terms of latency but the DF provides higher reliability. On the other hand, to enhance reliability under unknown channel conditions, rateless (fountain) codes~\cite{Byers98,Byers02,maymounkov2002online,Mackay}, such as Luby transform (LT) codes~\cite{Luby2002} and Raptor codes~\cite{Shokrollahi04Raptor,Shokrollahi06Raptor}, can be deployed. Rateless codes are capacity achieving for large block lengths, and are widely considered for multi-hop relay systems~\cite{Castura_TWC_2007,Gummadi_ITwkshp_2008}. In contradistinction with fixed-rate codes, rateless codes do not assume a pre-defined coding rate. Instead, the encoder potentially generates an unlimited number of codewords until decoding is successful. 
 
 Rateless codes are particularly beneficial for broadcasting/multicasting applications where a large number of receivers (e.g., massive machine-type and IoT communications) hinder channel estimation that can cause the feedback explosion problem. Yet, the problem with traditional fountain codes is that they show an \emph{all-or-nothing} decoding property such that the decoder only recovers a small portion of source messages until the very end. This performance is not desirable for real-time and delay-sensitive data delivery such as the scalable video streaming where the video quality is improved progressively as data arrives. 
 
 In order to improve the intermediate performance metric (i.e., number of recovered symbols as coding proceeds), there has been an extensive amount of research such as~\cite{Talari2010,Talari2012,Thomos2013,Cassuto2015,Hashemi2014,Hashemi2016,Jun2015,Jun2017}. These works mostly rely on either \emph{feedback channels} or encoding/decoding operations with \emph{higher complexities}, none of which is favorable to UDNs since: (i) due to the massive number of devices, leveraging the feedback channel may not be feasible, and (ii) low-power SBS, IoT, and sensory devices are not well-suited for high complexity encoding and decoding operations. Recently, memory-based LT encoders (MBLTEs)~\cite{Hayajneh2014,Hayajneh2015,shang2016second,shang2018,MBLTE_BEC} were proposed for improving the bit-error-rate (BER)/frame-error-rate (FER) performance of LT codes with relatively short block-lengths at the cost of adding memory into the encoder while maintaining the same low encoding/decoding complexity as LT codes.

 \emph{Our work is motivated by the UDN use-cases that call for delay-efficient data delivery with little-to-no coding feedback overhead and low-complexity encoding and decoding operations.} To this end, we first demonstrate that memory-based LT codes improve the intermediate performance without utilizing feedback signals.  Thus, they are especially suitable for UDN scenarios where feedback explosions and power/complexity are more of a concern than the amount of buffers/memory. Next, we propose a new relaying strategy based on DF for the SBSs in the UDN such that the real-time delay of data delivery to the users can be significantly improved. In addition, we propose an amicable rateless coding scheme that further improves the intermediate performance at the cost of adding temporary buffers. We present simulation results to verify that our proposed relaying strategy and the coding scheme are efficient in improving user experience in terms of real-time data delivery in UDNs.

The rest of the paper is organized as follows. Section~\ref{sec:SystemModel} describes the system model including the two-hop relay system, LT codes, and second-order MBLTEs. Section~\ref{sec: RelayStrategy} describes the new relaying strategy proposed for the small cells in the UDN. Section~\ref{sec:OrderedTransmission} describes the amicable encoding scheme for second-order MBLTEs. Section~\ref{sec:Results} presents simulation results, and Section~\ref{sec:Conclusions} concludes the paper.

\section{System Model}
\label{sec:SystemModel}
\subsection{Two-Hop Relay System}

\begin{figure}[t]
\centering\includegraphics[width=3in]{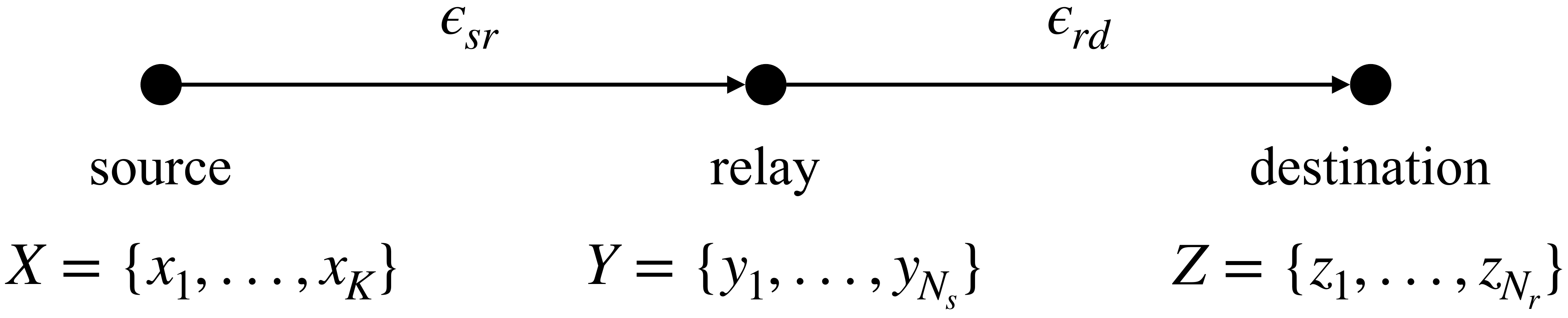}
\caption{The separated two-hop relay system. $\epsilon_{sr}$ and $\epsilon_{rd}$ denotes the erasure probability for the source-relay and the relay-destination link, respectively. The source node encodes $K$ input symbols $x_{1}$, \ldots, $x_{K}$ to $N_{s}$ output symbols $y_{1}$, \ldots, $y_{N_{s}}$; the relay node collects these output symbols then decodes, it then re-encodes the $K$ input symbols to $N_{r}$ output symbols $z_{1}$, \ldots, $z_{N_r}$ and forwards them to the destination; the destination node collects the output symbols then decodes.}
\label{fig:2HopModel}
\end{figure}

In this paper, we consider the three-node two-hop wireless relay system for lossy packet networks as shown in Fig. 2. The system constitutes a source, a relay, and a destination. Two independent links, the source-relay and the relay-destination, are considered in the system. We do not consider the direct link where the source communicates directly with the destination. We term this setting \textit{separated two-hop relay}, which is the simplest multi-hop communication problem and is frequently encountered in practice; for instance, in UDN scenarios, the source represents the MBS, the relay represents the SBS, and the destination represents the user. 

We assume binary erasure channels (BECs) for the links. The packet/symbol that goes through a BEC is either received correctly or erased by the channel. Let $X$ and $Y$ denotes the channel input and output, respectively, then $X \in \{0, 1\}$ and $Y \in \{0, 1, e\}$, where $e$ indicates an erasure, which occurs with probability $\epsilon$. 

For the traditional decode-and-forward (DF) relaying strategy, the communication is divided into two stages. In the first stage, the source encodes and transmits codewords to the relay. The relay starts decoding when it receives all of the codewords. It sends back an acknowledgement (ACK) to the source to terminate the transmission when decoding is successful. In the second stage, the relay re-encodes and transmits the codewords to the destination. The destination starts decoding when all codewords are received, and sends back an ACK to the relay in the event of a successful decoding. The DF achieves Shannon capacity of the separated two-hop relay system when the codeword length tends to infinity~\cite{Cover_IT_1979,Wang_Allerton_2017}. However, DF is significantly suboptimal for finite block-length codes. 


\subsection{LT Codes}
\textbf{Encoding:} For binary source data of length $K$, a $K$-by-$N$ generator matrix encodes the $K$ input symbols $ x_{1}, x_{2},\ldots, x_{K}$ to $N$ output symbols $y_{1}$, $y_{2}$,\ldots, $y_{N}$. The operation is performed with bitwise, modulo-2 additions, i.e.:
\begin{equation}
y_{n} = \sum_{k=1}^{K}x_{k}G_{kn},
\end{equation}
where $G_{kn}$ is an element of the generator matrix. The generator matrix $\mathbf{G}$ is obtained column by column. Each column represents a relationship between the input symbols and an output symbol. To obtain the output symbols, we sample the output symbol distribution $\Omega$ a total of $N$ times. The process for each sample is summarized below~\cite{Mackay}:
\begin{enumerate}

\item
Randomly choose the degree $d$ from a degree distribution $\Omega$.

\item
Uniformly choose $d$ rows at random in the $n$-th column in which to place a value of ``1".

\end{enumerate}

\textbf{Decoding:} The peeling decoder is used to decode LT codes over the BEC. Assume the decoder receives all $N$ output symbols. The decoding process is summarized below~\cite{Mackay}:

\begin{enumerate}

\item
Find an output symbol $y_{n}$ that is connected to only one input symbol $x_{k}$. If there is no such output symbol, the decoding process stops. If there is, go to Step 2.

\item
Set the input symbol equal to the output symbol: $x_{k}$=$y_{n}$.

\item
Find other output symbols $y_{i}$ that are connected to $x_{k}$.

\item
Add $x_{k}$ to each $y_{i}$ modulo~2.

\item
Remove all the edges connected to $x_{k}$.

\item
Return to Step 1.

\end{enumerate}
Next, we describe the memory-based LT encoding (MBLTE) scheme. 

\subsection{Second-Order MBLTE}
As outlined above, the LT encoder works by first sampling the degree distribution $\Omega$ to obtain a degree of $d$, and then sampling the input symbols a total of $d$ times. For a memoryless LT encoder, the current value of $d$ is all that is needed to proceed with sampling the input symbols. For a MBLTE, certain previous outcomes of $d$ are remembered which results in the input symbols being sampled differently. The order of the memory refers to the number of outcome types it remembers. For instance, remembering $i$ outcome types is called a $i^{th}$-order MBLTE. For the sake of exposition, we only consider the second-order MBLTE in this paper.  In the remainder of this paper, the term \textit{MBLTE} represents the second-order MBLTE unless otherwise stated.

The MBLTE remembers two outcomes with values $d = 1$ and $d = 2$. When $d = 1$, it samples the input symbol with the highest instantaneous degree, which is defined in~\cite{Hayajneh2014,Hayajneh2015} as the highest degree at the time when the current output symbol is being constructed. When $d = 2$, it samples one input symbol from those with the highest instantaneous degrees, then another one from those symbols with the second highest instantaneous degree. The details are shown in Algorithm~\ref{alg:MemoryOrder2}~\cite{shang2016second}. 

\begin{algorithm}[t]
\caption{The MBLTE encoding process.}
\label{alg:MemoryOrder2}
\begin{algorithmic}[1]
\State Initialize an empty set $S_{1}$
\For{$n = 1 : N$ }
\State{sample $\Omega$ to get $d$}
\If{$d= 1$}
\State{select the input symbol with the highest instantan-}
\State{eous degree without replacement, then put this}
\State{input symbol into set $S_{1}$}
\ElsIf {$d = 2$}
\State{first select an input symbol from the set $S_{1}$ unifor-}
\State{mly at random with replacement; then select an in-}
\State{put symbol with the highest instantaneous degree}
\State{except for those in set $S_{1}$ without replacement}
\Else
\For{$kk = 1 : d$}
\State{select an input symbol uniformly at random}
\State{without replacement}
\EndFor
\EndIf
\EndFor
\end{algorithmic}
\end{algorithm}

\section{Proposed Relaying Strategy}
\label{sec: RelayStrategy}
\begin{figure}
\centering\includegraphics[width=3.5in]{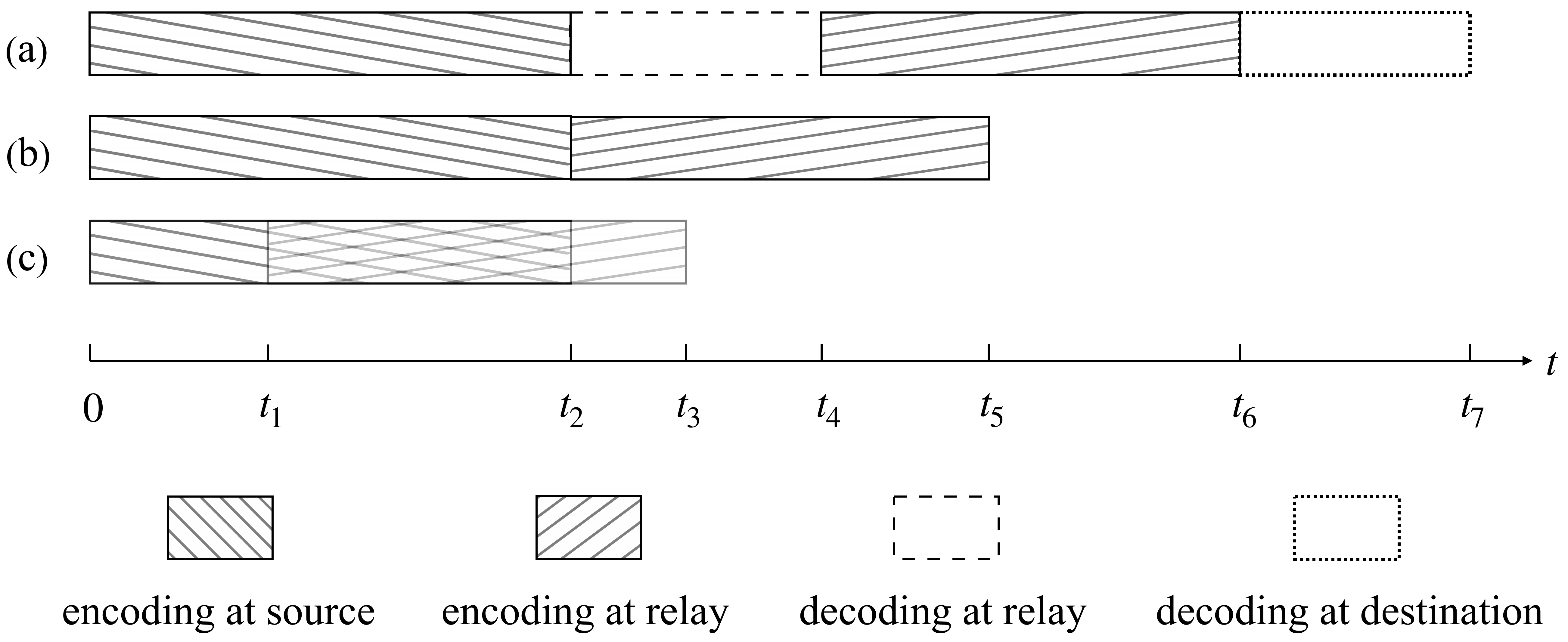}
\caption{ Motivation of the proposed relaying strategy. The communications between source and destination via relay is measured by the length of time. The box filled with left- and right-tilted stripes represents the time it takes for encoding at the source and the relay, respectively; the blank box with dashed and dotted borders represents the time it takes for decoding at the relay and the destination, respectively. }
\label{fig:CompStrategies}
\end{figure}

The DF relaying strategy provides reliable communication at the cost of non-negligible latency at the destination that grows linearly with the number of hops in the multi-hop system. The large latency at the destination results from the fact that the relay does not start re-encoding until the decoding is done. This problem would be greatly alleviated if decoding and re-encoding could operate simultaneously. Fig.~\ref{fig:CompStrategies} illustrates the motivation for proposing such a new relaying strategy. In the figure we assume zero propagation delay, i.e., once a codeword is generated by the encoder, it is received immediately by the decoder. In this way, the time it takes from source to destination is measured by the length of time it takes for the encodings and decodings in the system. Assume that the decoders at the relay and the destination are able to recover the source data after receiving $t_2$ and $t_6 - t_4$ time slots' coded data, respectively, as shown in (a) in Fig.~\ref{fig:CompStrategies}. We see that the latency at the destination can be shortened by replacing fixed-rate codes with rateless codes, where decoding is complete right after sufficient data is collected, as shown in (b). The latency can be further improved if the relay starts re-encoding/transmitting before decoding is complete as shown in (c). 

To have the relay decode and re-encode simultaneously, two requirements must be met: (\romannumeral 1) the relay node operates in the full-duplex mode; and (\romannumeral 2) the relay recovers source data as coded data arrives. Nowadays as the usage of full-duplex relays is widespread, and as techniques such as~\cite{Riihonen2011_TSP,Atzeni_ICC_2016,Kabir_Globecom_2017} that mitigate the self-interference of full-duplex relays are well under development, the first requirement is no longer a concern. To meet the second requirement, rateless codes with good intermediate performance are needed. Now that we have discussed in Section~\ref{sec:Intro} that MBLTE is a good candidate in this regard, the second requirement is also met.


\subsection{Partial Decode-and-Forward Relaying}
In this section, we consider full-duplex relays with rateless codes in the separated two-hop system, and assume no self-interference. Discussions on weighing between half- or full-duplex relays or the self-interference of full-duplex relays are beyond the scope of this paper, and can be found in the literature, cf. e.g.~\cite{Aggarwal2012_ITwkshp,Riihonen2011_TWC,Duarte2012_TWC,Aryafar2012,Razlighi_Globecom_2017}. We also make the following assumptions: (\romannumeral 1) each output symbol takes one time slot to be generated by the encoder; (\romannumeral 2) the output symbol is immediately received and processed by the decoder once it is generated; and (\romannumeral  3) the decoder can recover as many input symbols as possible in one time slot. We let $\mathcal{R}(t)$ and $\mathcal{D}(t)$ denote the set of recovered input symbols by the decoder at the relay and the destination, respectively, at time $t$, thus $\mathcal{R}(t)$ and $\mathcal{D}(t)$ are empty initially. We propose a partial DF (PDF) relaying strategy, where the relay starts re-encoding immediately once $\mathcal{R}(t)$ becomes non-empty. The relay encodes over the input symbols in $\mathcal{R}(t)$ dynamically until it is notified by the destination to terminate. We assume that the relay starts encoding in the same time slot as $\mathcal{R}(t)$ becomes non-empty.  

\begin{figure}
\centering\includegraphics[width=3.5in]{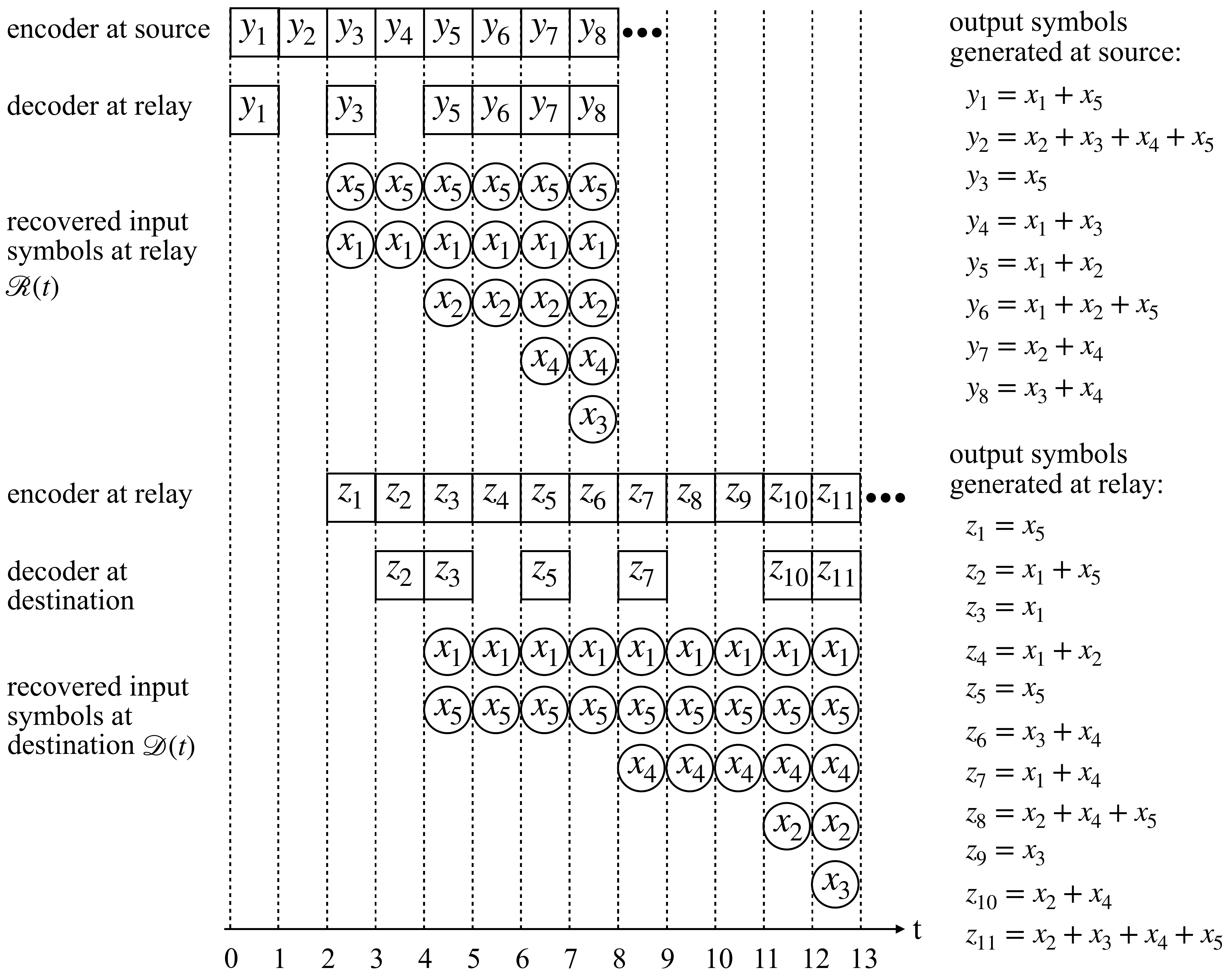}
\caption{Example of the proposed PDF relaying strategy. }
\label{fig:NewStrategy}
\end{figure}

To better illustrate our proposed PDF relaying strategy, we resort to an example shown in Fig.~\ref{fig:NewStrategy}. The circles and rectangles represent input and output symbols, respectively.  There are five input symbols ${x_1, \ldots, x_5}$. The encoder at the source encodes these symbols into a potentially unlimited number of output symbols $y_i$ using  modulo-2 operations. The decoder at the relay process each successfully-received encoded symbol ${y_i}$. For instance, we assume that $y_2$ and $y_4$ are erased by the channel. The relay recovers $x_5$ and $x_1$ when $2 < t \leq 3$ and $\mathcal{R}$(t) becomes non-empty as shown in \eqref{eq:R(t)}. The relay immediately starts re-encoding by encoding over $x_5$ and $x_1$ and generates its first output $z_1$, which is transmitted to the destination but is erased by the channel. The relay generates an output symbol by re-encoding over the input symbols in $\mathcal{R}$(t) during each time slot thereafter until $t = 13$ where the destination successfully completes decoding. The dynamic evolution of the sets $\mathcal{R}(t)$ and $\mathcal{D}(t)$ is shown in~\eqref{eq:R(t)} and~\eqref{eq:D(t)}, respectively. 
\begin{equation}
\label{eq:R(t)}
\mathcal{R}(t) = 
\begin{cases}
\{\},                                  & 0 \leq t \leq 2\\
\{x_5, x_1\},                     & 2 < t \leq 4\\
\{x_5, x_1, x_2\},               & 4 < t \leq 6\\
\{x_5, x_1, x_2, x_4\},        & 6 < t \leq 7\\
\{x_5, x_1, x_2, x_4, x_3\}, & t > 7\\
\end{cases}
\end{equation}

\begin{equation}
\label{eq:D(t)}
\mathcal{D}(t) = 
\begin{cases}
\{\},                       & t \leq 4\\
\{x_1, x_5\},               & 4 < t \leq 8\\
\{x_1, x_5, x_4\},           & 8 < t \leq 11\\
\{x_1, x_5, x_4, x_2\},      & 11 < t \leq 12\\
\{x_1, x_5, x_4, x_2, x_3\}, & t > 12.\\
\end{cases}
\end{equation}

{Next, we propose \emph{amicable} encoding scheme that will be used in conjunction with MBLTE.}

\section{Amicable Encoding Scheme}
\label{sec:OrderedTransmission}
Based on the proposed relaying strategy, the relay starts re-encoding and transmitting immediately once an input symbol is recovered. The earlier the relay starts re-encoding, the earlier the destination starts reception and decoding; thus, the real-time delay experienced by the users is reduced. As such, in this section we aim to propose an amicable MBLTE so that (\romannumeral 1) it starts recovering input symbols earlier; and (\romannumeral 2) it recovers more input symbols during each time slot.

For the MBLTE, the decoder is not able to recover any input symbols until a degree-$1$ output symbol is received. Therefore, a good encoder should generate and transmit degree-$1$ output symbols early. In addition, according to the encoding process of the MBLTE, input symbols that are connected to the degree-$1$ output symbols are highly likely to be also connected to the degree-$2$ output symbols. Once the above-mentioned input symbols are recovered, their neighboring degree-$2$ output symbols reduce to degree-$1$, enabling the decoder to proceed in terms of recovering more input symbols. Therefore, a good MBLTE should also transmit degree-$2$ output symbols early, following right after the transmission of the degree-$1$ symbols. However, because the encoder does not know how many output symbols to generate in the rateless scenario, it is also unknown how many degree-$1$ or degree-$2$ output symbols will be generated, not to mention ordering the transmissions.

To solve this problem, we propose an amicable rateless encoding scheme that divides the encoding process into two stages where the first stage can be viewed as a fixed-rate transmission and the second stage is rateless. We now predetermine the number of output symbols $N$ for the first stage. According to~\cite{MBLTE_BEC}, the optimal memory order is the smallest integer $i$ that satisfies
\begin{equation}
\label{eq: OptOrder}
N\sum \textstyle_{d = 1}^{i} \Omega_{d} \geq K,
\end{equation}
where $\Omega_d$ is the probability that an output symbol is degree-$d$. In our case where the memory order is $i = 2$, we rearrange the inequality as
\begin{equation}
\label{eq: determineN}
N \geq K / (\Omega_1 + \Omega_2).
\end{equation}
Although~\eqref{eq: determineN} does not apply for the rateless scenario where $N$ is unknown, it is a useful tool in determining the number of output symbols for the first stage of encoding. According to~\cite{MBLTE_BEC}, ideally without channel erasures the decoder is able to recover all input symbols with the reception of $N$ output symbols if $N$ satisfies~\eqref{eq: determineN}, while the performance degrades as $N$ grows if the channel erasure probability is nonzero. Therefore, selecting $N$ to be the smallest integer that satisfies~\eqref{eq: determineN} results in the second-order MBLTE being optimal. We thereby modify the MBLTE as detailed in Algorithm~\ref{alg:amicableMBLTE}.

\begin{algorithm}
\caption{The encoding process of the amicable MBLTE.}
\label{alg:amicableMBLTE}
\begin{algorithmic}[1]
\State Initialize two empty sets: $S_{1}$ and $S_d$
\State Define $N$ as the smallest integer that satisfies~\eqref{eq: determineN}
\State {Sample $\Omega$ a total of $N$ times, store values in $S_d$. Rearrange the elements in $S_d$ such that they are in ascending order}
\For{$n = 1 : \infty$ }
\If{$n \leq N$}
\State{$d = $ the first element in $S_d$}
\State{remove the first element from $S_d$}
\Else
\State{sample $\Omega$ to get $d$}
\EndIf
\State{Select $d$ input symbols as detailed in Alg.~\ref{alg:MemoryOrder2}, Lines 4--18}
\EndFor
\end{algorithmic}
\end{algorithm}

Because the amicable MBLTE does not change the degree distribution of the output symbols, and because it does not affect the decoding process, it maintains the same low encoding and decoding complexities as the traditional MBLTE. Compared to the traditional one, the amicable encoder requires $N$ additional buffers at the initial stage of encoding. However, $N$ is not large and only depends on the input symbol length $K$ when the degree distribution is fixed regardless of the channel condition. Moreover, the number of buffers decreases linearly as encoder proceeds as detailed in Algorithm~\ref{alg:amicableMBLTE}. For applications where buffer size is less of a concern than delay or complexity, the amicable MBLTE is beneficial.

\section{Simulation Results}
\label{sec:Results}
In this section, we present simulation results to evaluate the proposed partial decode-and-forward (PDF) strategy. Because the PDF would benefit from rateless codes that have good intermediate performance, we first compare the traditional MBLTE with the amicable MBLTE encoder and show that the latter has a better intermediate performance than the former. We then employ the amicable codes as the rateless codes in the two-hop system to evaluate the performance of our proposed PDF method against the decode-and-forward (DF) scheme. In the simulations, we fix the length of the input symbols at $K = 256$, and adopt the robust soliton distribution (RSD) with $c = 0.03$ and $\delta = 0.5$~\cite{Luby2002}.

\begin{figure}[t]
\centering\includegraphics[width=3.5in]{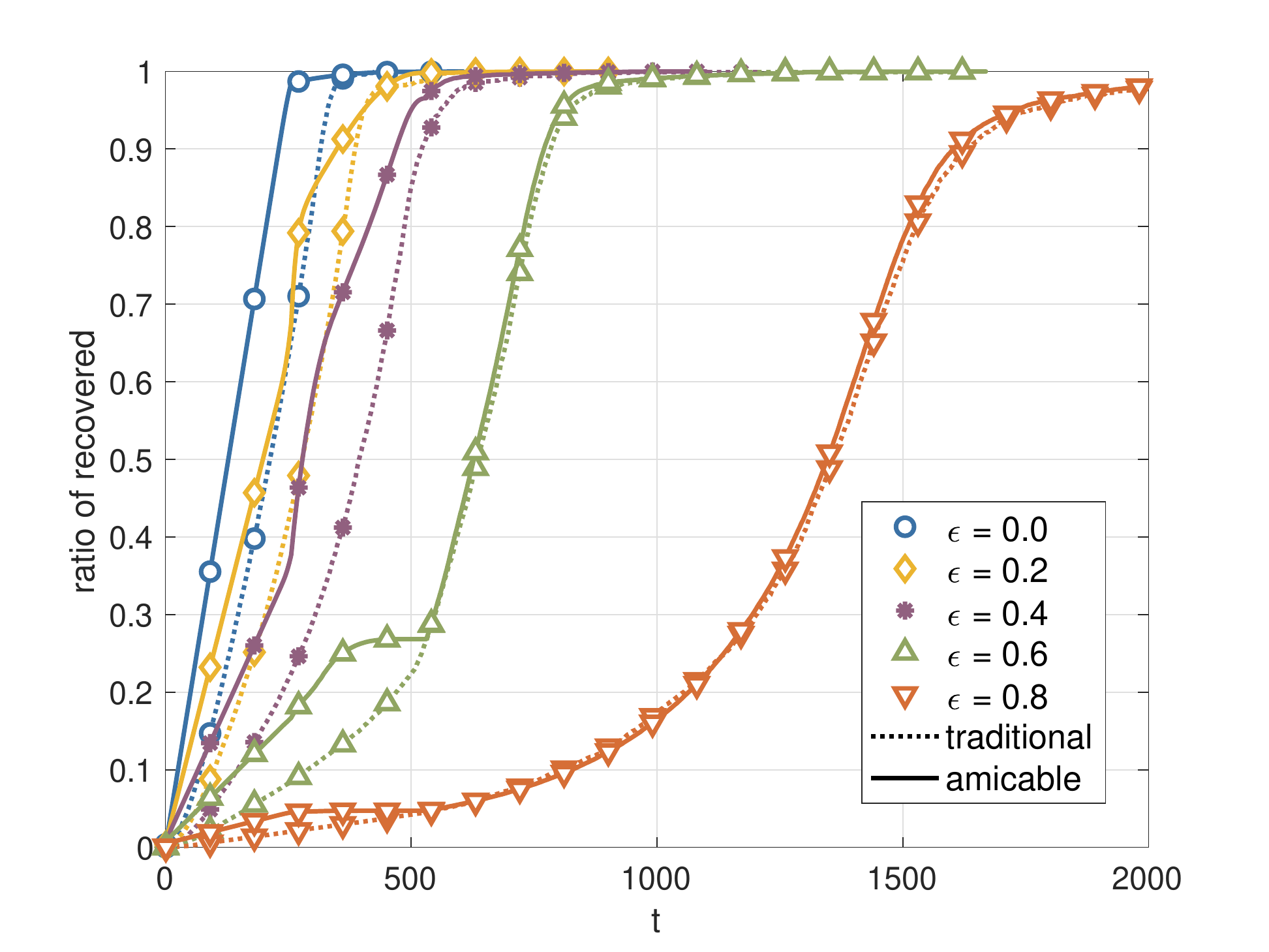}
\caption{Intermediate performance comparison of traditional MBLTE and amicable MBLTE as coding proceeds. The dotted lines represent the traditional encoder and the solid lines represent the amicable encoder. The channel erasure probability is denoted by $\epsilon$. 
	The input symbol length is fixed at $K = 256$.}
\label{fig:K256c003_MBLTE_VS_modified1}
\end{figure}

\subsection{Performance of Amicable MBLTE for Single-hop}
Fig.~\ref{fig:K256c003_MBLTE_VS_modified1} shows the intermediate performance of the {traditional MBLTE} and amicable MBLTE under various erasure probabilities ranging from $\epsilon = 0$ to $\epsilon = 0.8$. From the results, we note that although the performance of each encoder degrades as channel conditions worsen, the amicable encoder outperforms the traditional encoder for all channel conditions shown in the figure. Specifically, the gap between the encoders is about the same for the cases where $\epsilon = 0$, $\epsilon = 0.2$, and $\epsilon = 0.4$; the gap becomes negligible during most of the decoding period when the channel erasure probability increases to $\epsilon = 0.6$ and $\epsilon = 0.8$. One of the interesting phenomena in the figure is that for $\epsilon = 0.6$ there is an obvious performance gain of the amicable encoder relative to the traditional one at the early stage of decoding where $t \leq 500$. It should be noted that we predetermine the number of output symbols $N$ for the first stage of encoding in Algorithm~\ref{alg:amicableMBLTE} such that the amicable encoder differs from the traditional one only in the first $N$ output symbols. For the parameters used in this simulation, the predetermined $N$ is {around} $500$. Therefore, this phenomenon appearing in the case of $\epsilon = 0.6$ is consistent with the amicable encoding process. Moreover, it verifies that the proposed amicable encoder is effective in improving the intermediate performance of the traditional MBLTE. A similar phenomenon is also observed for the case of $\epsilon = 0.8$ although it is less obvious because of the poor channel condition. This is not observed for cases where $\epsilon \leq 0.4$ because both codes have recovered most of the input symbols and are close to completion of decoding at $t = 500$.

\begin{figure}[t]
\centering\includegraphics[width=3.5in, trim=0cm 0cm 0cm 0.8cm, clip]{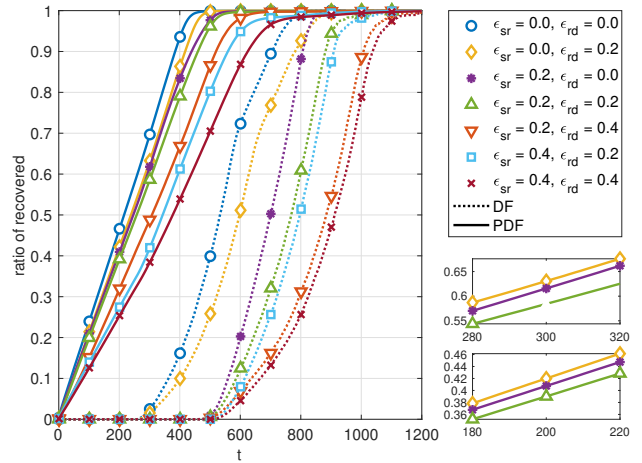}
\caption{Comparison of DF and PDF using the amicable MBLTE as coding proceeds. The dotted lines represent DF and the solid lines represent PDF. The vertical axis represents the ratio of recovered input symbols at the destination. The input symbol length is fixed at $K = 256$. Various respective channel erasure probabilities $\epsilon_{sr}$ and $\epsilon_{rd}$ for the source--relay and relay--destination links are evaluated. The two sub-figures give a detailed view of PDF for some channel conditions around $t = 200$ and $t = 300$.}
\label{fig:K256c003modified1_DFvsPDF_delay}
\end{figure}

\subsection{Performance of Partial Decode-and-Forward for Two-hop}
Fig.~\ref{fig:K256c003modified1_DFvsPDF_delay} compares the performance of PDF with DF by employing the amicable MBLTE as the rateless codes at the source and the traditional MBLTE at the relay. Various channel erasure probabilities are evaluated with $\epsilon_{sr}$ (source to relay) and $\epsilon_{rd}$ (relay to destination) ranging from $0$ to $0.4$, respectively. From the results, we see that PDF significantly improves the performance of DF for all channel conditions. Specifically, for each case with the PDF strategy, the destination starts to recover input symbols right from the beginning; while with DF, the destination is not able to recover any input symbols until $t = 256$. In addition, in each erasure scenario shown in the figure, DF only recovers less than 20\% of input symbols at the time when 97\% are recovered by PDF.

Moreover, from the figure we also notice that the gap between the best and worst scenarios is much smaller for PDF than it is for DF, i.e. PDF recovers 50\% of the input symbols when $t = 216$ in the case of $\epsilon_{sr} = 0$ and $\epsilon_{rd} = 0$, and when $t = 377$ in the case of $\epsilon_{sr} = 0.4$ and $\epsilon_{rd} = 0.4$, resulting in a gap of $\Delta t = 161$; while PDF does so at $t = 533$ and $t = 912$, resulting in a gap of $\Delta t = 379$, which is more than double that of PDF. Therefore, PDF is much more robust against performance degradation due to erasures.

\section{Conclusions}
\label{sec:Conclusions}
In this paper, we have proposed a partial decode-and-forward (PDF) relaying strategy and have realized it with rateless codes. We have argued that the proposed PDF benefits from rateless codes that have good intermediate performance. We have showed that the recently developed MBLTE has good intermediate performance and we have proposed an amicable encoding scheme for a further performance gain at the cost of additional temporary buffers. We have presented simulation results to show that our amicable encoding algorithm is efficient in further improving the intermediate performance especially when the channel erasure probability is not too high, and that our proposed PDF not only significantly improves the performance of DF but is also more robust against performance degradation due to erasures. Overall, the proposed encoding scheme and the relaying strategy are efficient towards near real-time data delivery in the UDN and also the Internet of Things (IoT) scenario where many devices/sensors are power-limited thus a direct link from the source to destination is impossible. In our future work, we will investigate the tradeoffs between the delay and the overhead of the PDF strategy and extend the work to scenarios with more than two hops.


\bibliographystyle{ieeetr}
\bibliography{ID}
\end{document}